\documentclass{elsart}
\usepackage{color}
\usepackage{graphicx}
\usepackage{amsmath,amssymb}

\def\epjc#1#2#3{Eur.\ Phys.\ Jour.\ C{\bf #1}, #2 (#3)}
\def\ibid#1#2#3{{\it ibid.} {\bf #1}, #2 (#3)}

\def\jhep#1#2#3{Jour. High Energy Phys. {\bf #1}, #2 (#3)}

\def\plb#1#2#3{Phys. Lett. B {\bf #1}, #2 (#3)}

\def\prd#1#2#3{Phys. Rev. D {\bf #1}, #2 (#3)}
\def\prl#1#2#3{Phys. Rev. Lett. {\bf #1}, #2 (#3)}

\begin{document}

\begin{frontmatter}
\title{Hadron Production in Ultra-relativistic
Nuclear Collisions: Quarkyonic Matter and a Triple Point
in the Phase Diagram of QCD}
\author[gsi]{A.\ Andronic},
\author[wroclaw,dubna]{D.\ Blaschke},
\author[gsi,emmi,tud,fias]{P.\ Braun-Munzinger},
\author[capetown]{J.\ Cleymans},
\author[yukawa]{K.\ Fukushima},
\author[bnl,riken]{L.D.\ McLerran},
\author[tud]{H.\ Oeschler},
\author[bnl]{R.D.\ Pisarski},
\author[gsi,wroclaw,bielefeld]{K.\ Redlich},
\author[fias,munich]{C.\ Sasaki},
\author[bielefeld]{H.\ Satz},
and \author[heidelberg] {J.\ Stachel}

\address[gsi]{GSI Helmholtzzentrum f\"ur Schwerionenforschung,
  D-64291 Darmstadt, Germany}
\address[wroclaw]{Institute for Theoretical Physics, University of
  Wroclaw, 50-204 Wroclaw, Poland}
\address[dubna]{Bogoliubov Lab.\ for Theoretical Physics, JINR Dubna,
  141980 Dubna, Russia}
\address[emmi]{ExtreMe Matter Institute EMMI, GSI, D-64291 Darmstadt,
  Germany}
\address[tud]{Technical University Darmstadt, D-64289 Darmstadt,
  Germany}
\address[fias]{Frankfurt Institute for Advanced Studies, J.W.\ Goethe
  University, D-60438 Frankfurt, Germany}
\address[capetown]{Physics Department, University of Cape Town, South
  Africa}
\address[yukawa]{Yukawa Institute for Theoretical Physics, Kyoto
  University, Kyoto, Japan}
\address[bnl]{Physics Dept., Brookhaven National Laboratory Upton,
  NY-11973, USA}
\address[riken]{RIKEN/BNL Research Center, Brookhaven National
  Laboratory Upton, NY-11973, USA}
\address[bielefeld]{Fakult\"at fur Physik, Universit\"at Bielefeld, D-33501 Bielefeld, Germany}
\address[munich]{Physik-Department, Technische Universit\"at
  M\"unchen, D-85747 Garching, Germany}
\address[heidelberg]{Physikalisches Institut der Universit\"at
  Heidelberg, D-69120 Heidelberg, Germany}

\begin{abstract} We argue that features of hadron production in relativistic
 nuclear collisions, mainly at
CERN-SPS energies, may be explained by the existence of three
forms of matter:  Hadronic Matter, Quarkyonic Matter, and a Quark-Gluon
Plasma.
We suggest that these
meet at a triple point in the QCD phase diagram. Some of the
features explained, both qualitatively and semi-quantitatively, include the
curve for the decoupling of chemical equilibrium,
along with the non-monotonic behavior of strange
particle multiplicity ratios at center of mass energies near $10$~GeV. If
the transition(s) between the three phases are merely crossover(s), the
triple point is only approximate.
\end{abstract}


\end{frontmatter}

\section{Introduction}

The SPS heavy ion program at CERN resulted in some of the first experimental
data on heavy ion collisions at ultrarelativistic energies, see, e.g.,
\cite{Specht:2001qe}.  A summary of these data and implications for the
possible formation of a new state of matter were  announced in a CERN press release
\cite{Heinz:2000bk}. In this paper we consider some generic features
discovered in heavy ion experiments at the SPS. This gives us a general
overview of how the collisions of heavy ions evolve in going from low
energies, as studied at the SIS (GSI) and the AGS (BNL), to higher energies,
at RHIC (BNL) and soon at the LHC (CERN) \cite{Stachel:1998rc,Arsene:2004fa,Back:2004je,Adams:2005dq,Adcox:2004mh,BraunMunzinger:2007zz}.
\begin{figure}[htbp]
\begin{center}
\resizebox*{!}{8.0cm}{\includegraphics{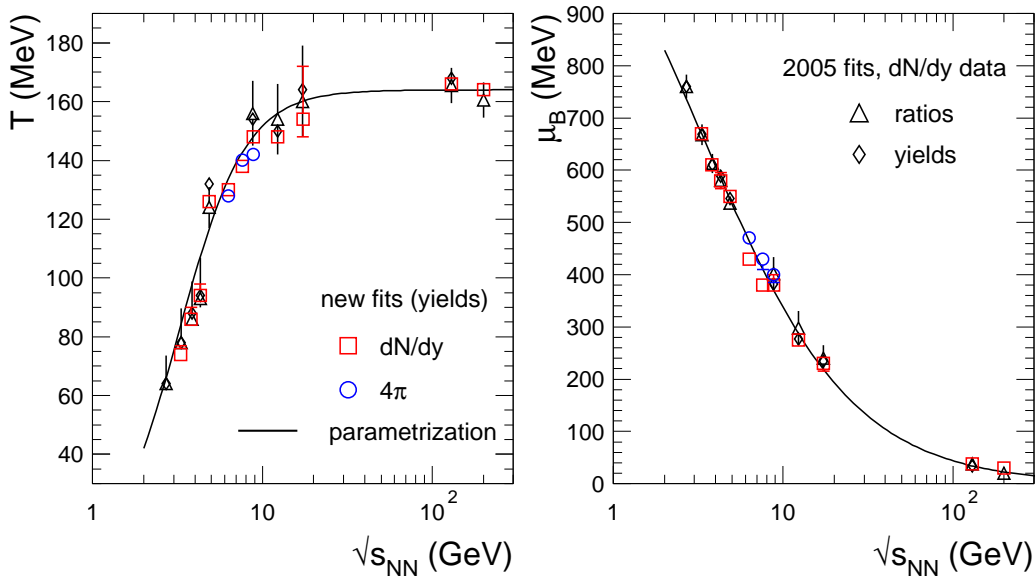}}
\end{center}
\caption{The temperature and baryon chemical potential of Statistical Model
 fits to hadro-chemical abundances as a function of center of mass energy per
 nucleon pair for collisions of heavy nuclei (Figure taken from
 ~\cite{Andronic:2008gu,Andronic:2009qf}).}
\label{energy_t_mu}
\end{figure}

In particular, we concentrate on hadron  abundances in heavy
ion collisions. These have been widely studied using resonance gas models.  By
assuming that the observed particle yields are generated at a common surface
at which all particles decouple, values of the baryon chemical potential,
$\mu_B$, and temperature, $T$, on this surface, can be extracted.  Fitting
these two parameters, $\mu_B$ and $T$, together with the  volume parameter gives values for the particle
abundances which are in close agreement with
experiment \cite{BraunMunzinger:1994xr,Cleymans:1999st,Heinz:1999kb,Letessier:2000ay,BraunMunzinger:2001ip,Florkowski:2001fp,Becattini:2005xt,Becattini_n,Cleymans:2005zx,Andronic:2005yp,hwa,Andronic:2008gu,Andronic:2009qf,corw}.
The resulting values of $\mu_B$ and $T$ are shown in Fig.~\ref{energy_t_mu} as
 functions of center-of-mass energy per nucleon pair.

We note that, near 10 GeV center of mass energy, the temperature saturates
with increasing beam energy, reaching an asymptotic value of about 160 MeV, while
the baryon chemical potential decreases smoothly.

\begin{figure}[htbp]
\begin{center}
\resizebox*{!}{10.0cm}{\includegraphics{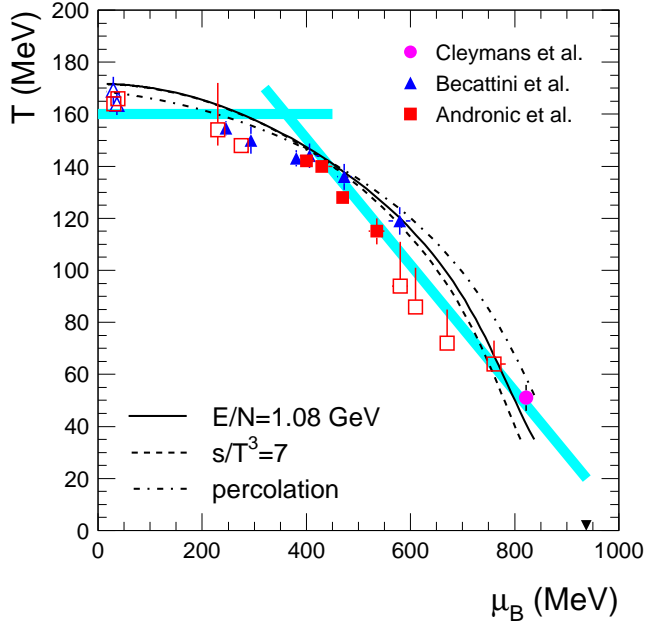}}
\end{center}
\caption{The decoupling temperatures and chemical potentials extracted by
Statistical Model fits to experimental data.
The freeze-out points are  from Refs.  \cite{Becattini:2005xt} 
and \cite{Andronic:2008gu,fr1,fr2}. The open points are obtained from fits 
to mid-rapidity whereas the full-points to $4\pi$ data.  The inverse 
triangle at $T=0$ indicates the position of normal nuclear matter.
The lines are different model calculations to quantify these points
\cite{corw,per,1gev}.
The shaded  lines are drawn to indicate different regimes in this diagram (see text).
}
\label{line}
\end{figure}

Plotting these temperature-chemical potential pairs for all available energies
results in a phase diagram-like picture as is illustrated in Fig. \ref{line}.
In the $\mu_B$ region from 800 to 400 MeV, as $T$ increases from 50 to 150
MeV, the experimental points rise approximately linearly. In contrast, below
$\mu_B \sim 400$~MeV, the temperature is approximately constant, $T \simeq
160$~MeV. The highest collision energies studied to date at RHIC are those for
which $\mu_B\sim 25$~MeV.  Also shown on this plot are lines of fixed energy per particle and fixed  entropy density per $T^3$; also shown is a line of  hadron percolation (see below).

These experimental results can be compared to phase transition points computed
on the lattice \cite{lattice_review,ghk}.  Numerical simulations in lattice
QCD can be performed at nonzero temperature, and small values of $\mu_B$
without running into problems of principle. At $\mu_B = 0$, these simulations
indicate that there is no true phase transition from Hadronic Matter to a
Quark-Gluon Plasma, but rather a very rapid rise in the energy density at a
temperature $T_c$ which lines  in $ 160-190$~MeV within the systematic errors.
Further, studies using the lattice technique imply that $T_c$ decreases very 
little as $\mu_B$ increases, at least for moderate  values of $\mu_B$.

\begin{figure}[htb]
\begin{tabular}{lr} \begin{minipage}{.49\textwidth}
\hspace{-.3cm}\includegraphics[width=1.\textwidth]{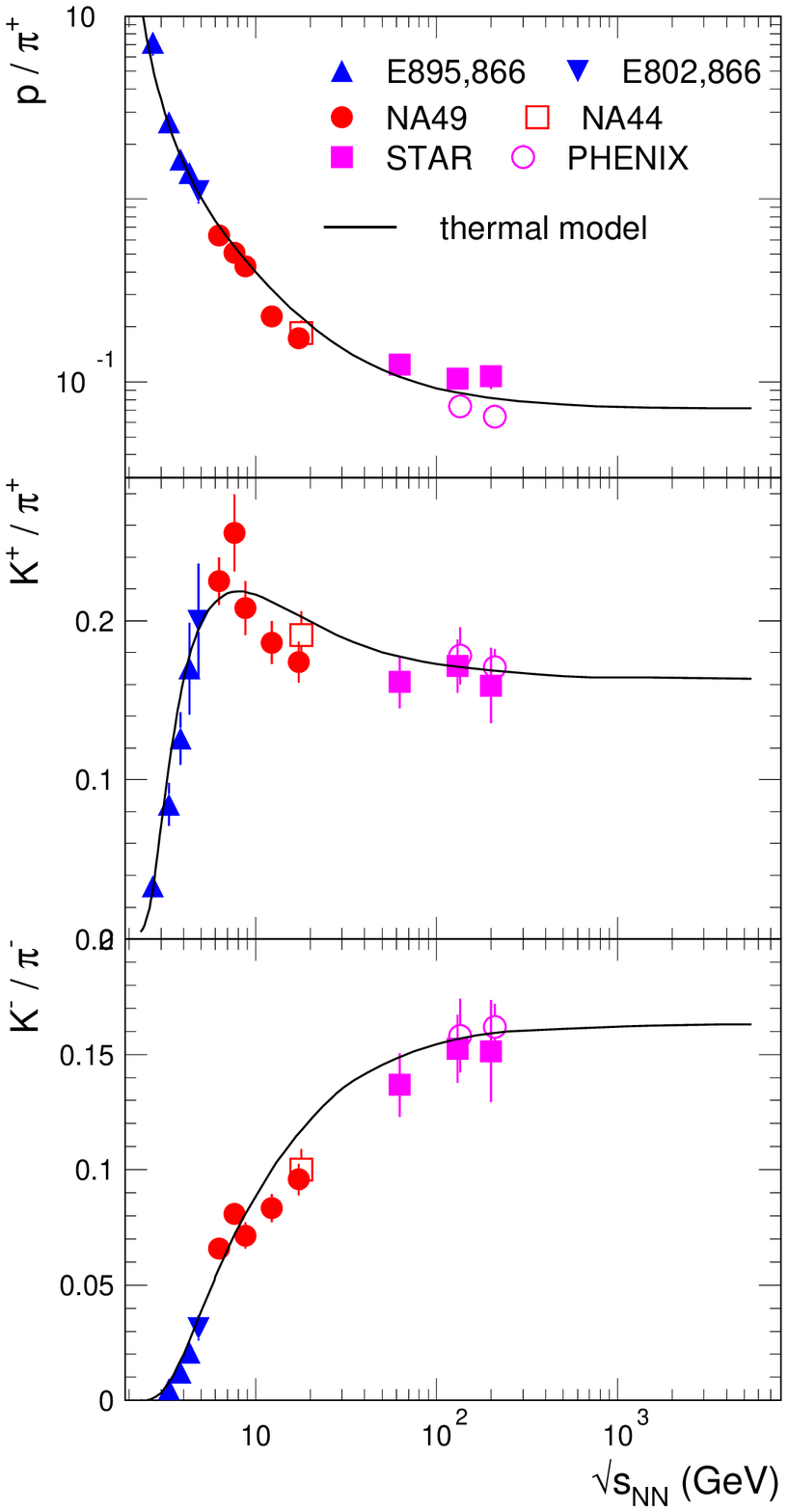}
\end{minipage} &\begin{minipage}{.49\textwidth}
\hspace{-.5cm}\includegraphics[width=1.\textwidth]{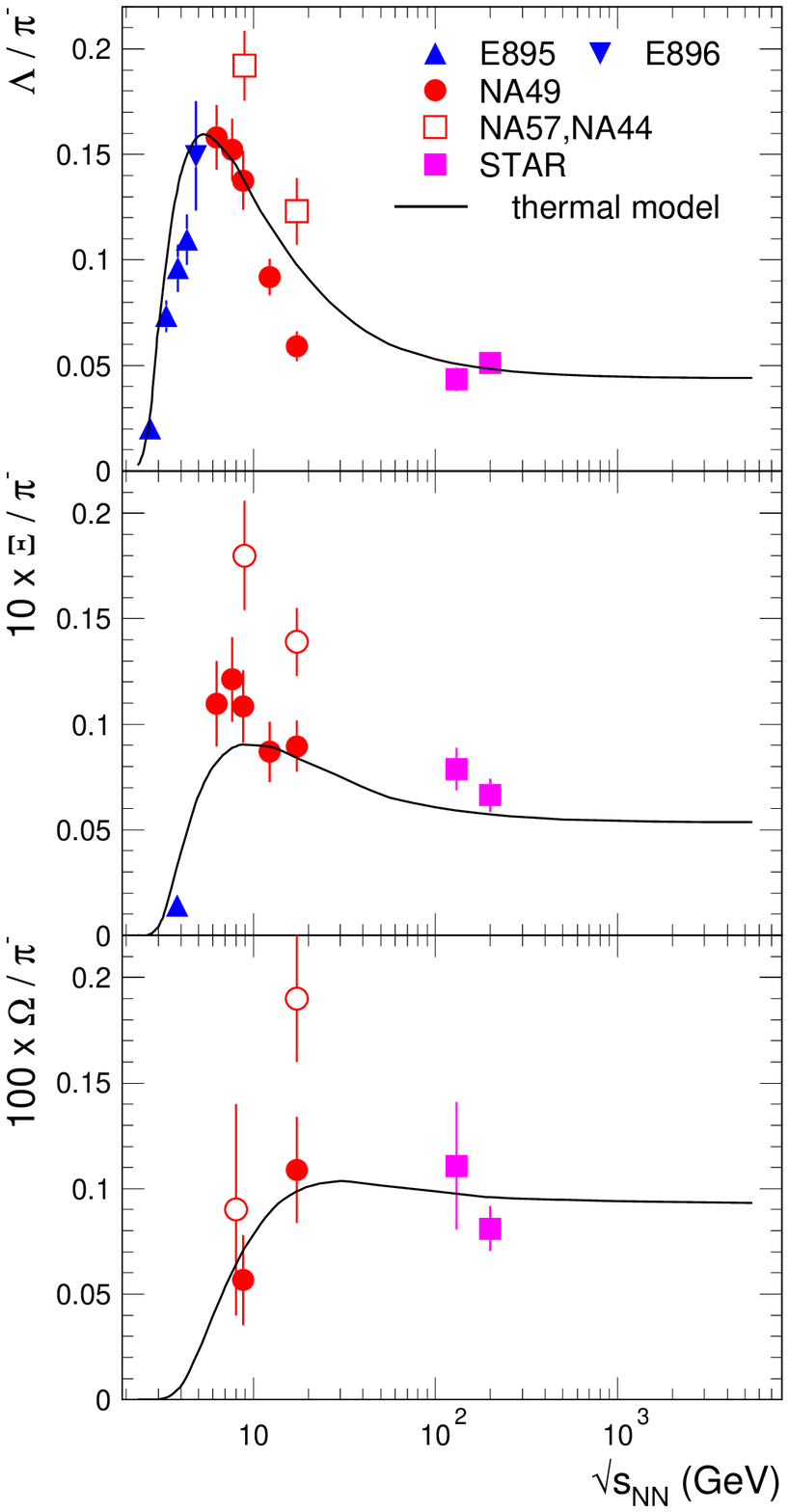}
\end{minipage} \end{tabular}
\caption{Energy dependence of hadron yields relative to pions. The points are experimental data from verious experiments.  Lines are results of the Statistical Model calculations.  The figure is taken from
 ~\cite{Andronic:2009qf,Andronic:2008gu}).}
\label{fig_k2pi}
\end{figure}

\begin{figure}[htbp]
\begin{center}
\resizebox*{!}{9.0cm}{\includegraphics{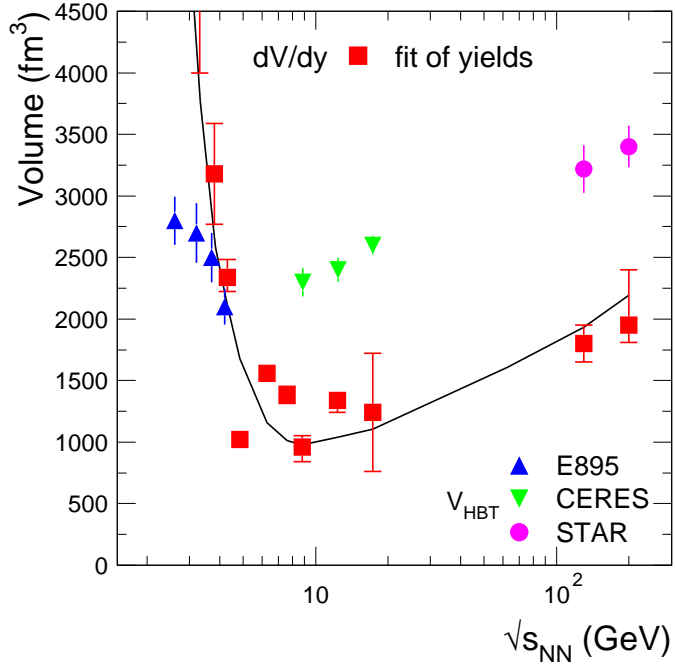}}
\end{center}
\caption{Energy dependence of the volume for central nucleus-nucleus collisions. The chemical
freeze-out volume  dV/dy  for one unit of rapidity (boxes) taken   from  Ref. \cite{Andronic:2005yp} is compared to the kinetic freeze-out volume V$_{\rm HBT}$ (filled circles and triangles)
 from   Ref. \cite{Adamova:2002ff}.
The line is the Statistical Model calculations with thermal parameters from Fig. \ref{energy_t_mu}.}
\label{size}
\end{figure}

With the  parametrizations of $T$ and $\mu_B$ from  Fig.~\ref{energy_t_mu} one can compute the energy
dependence of the production yields of various hadrons relative to pions,
shown in Fig.~\ref{fig_k2pi}.
Important for our purposes is the observation that there are peaks in the
abundances of strange to non-strange particles at center of mass energies near
10 GeV. In particular, the $K^+/\pi^+$  and $\Lambda/\pi$  ratios exhibit 
rather pronounced maxima there. 
We further note that in the region near 10 GeV, there is also a minimum in the
chemical freeze-out volume \cite{toneev,Andronic:2005yp} obtained from the 
Statistical Model fit to particle 
yields \cite{Andronic:2005yp,Andronic:2009qf},  
as well as in the volume obtained from the Hanbury-Brown and Twiss (HBT) 
radii of the fireball \cite{Adamova:2002ff}. 
The energy dependence of the volume parameters is shown in Fig. \ref{size}.

These experimental observations have long resisted interpretation in
terms of a transition between Hadronic Matter and a Quark-Gluon
Plasma\footnote{We note the interpretation given in \cite{Gazdzicki:1998vd},
obtained within a schematic 1st order phase transition model.}. 
The general structures observed in the data are
 well reproduced only by the most recent model calculations
\cite{Andronic:2008gu}.  There,  it is argued that these structures
arise due to the interplay between the limit in hadronic temperature (see
Fig.~\ref{energy_t_mu}) due to the QCD phase transition and the rapid decrease
of $\mu_B$ with increasing energy, thereby establishing a connection between
Hadron Gas and Quark-Gluon Plasma.
The possible existence of a  critical
endpoint is, however, not relevant for these considerations.

The above described structures seem puzzling if the corresponding energies
would probe a critical endpoint in the QCD phase diagram \cite{Stephanov}.  Near a
critical point, lighter particles, such as pions, should be affected more than
heavier particles, such as kaons; HBT radii should also increase.  Both of
these features are not easily linked to the trends in the data.

We will discuss the
relationship between the above  Statistical Model descriptions of the transition to both
the Quark-Gluon Plasma and Quarkyonic Matter, the triple point where three phases of matter
coexist, and the underlying contribution to the spectrum of strange particles below, and argue that generic features of these curves may be explained in this context.

\section{Quarkyonic Matter and the QCD Phase Diagram}

\begin{figure}[htbp]
\begin{center}
\resizebox*{!}{8.cm}{\includegraphics{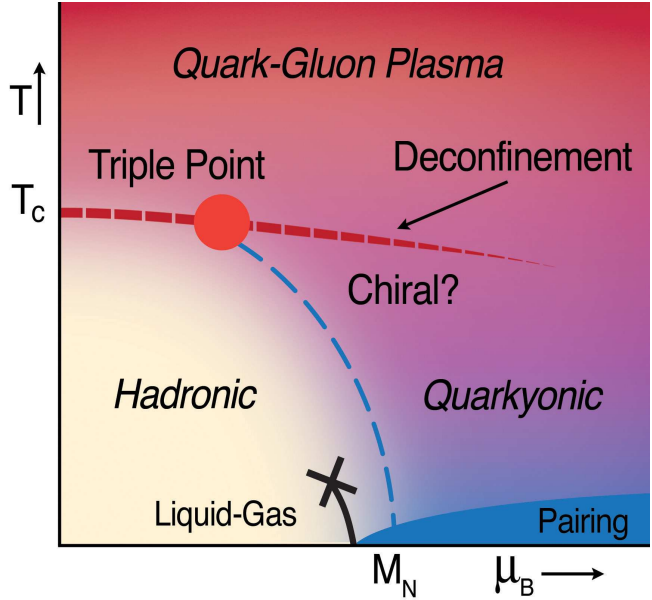} }
\end{center}
\caption{The phase diagram of strongly interacting matter.}
\label{Quarkyonic}
\end{figure}

In the following we show that by considering Quarkyonic Matter, which was
recently proposed
\cite{McLerran:2007qj,Hidaka:2008yy,McLerran:2008ua,Fukushima:2008wg,Glozman:2007tv},
the two regimes observed in the phase diagram and described above can be
understood as arising from a triple point where Hadronic Matter, the
Quark-Gluon Plasma, and Quarkyonic Matter all coexist.  This triple point is
located where the temperature is reaching its limiting value and, hence, is
naturally also situated in the vicinity of the peaks in the observed hadron
production ratios. A sketch of a possible phase diagram for QCD is shown
in Fig. \ref{Quarkyonic}.

There are hadrons in the lower, left-hand corner of this phase diagram, at low
temperatures and $\mu_B$.  There are two, qualitatively distinct, phase
boundaries by which one can leave Hadronic Matter.
  The first, is to
increase  the temperature at low $\mu_B$ until it is beyond $T_c$.  This is
the usual transition from a meson-dominated phase\footnote{We note that, 
at chemical freeze-out, the density of baryons and anti-baryons, $n_B$, 
is similar in this regime to that at large-$\mu_B$ ($n_B\simeq$0.12 fm$^{-3}$)
\cite{Andronic:2005yp}.} 
to a Quark-Gluon Plasma.  This phase boundary
is probed by collisions at high SPS energies, and by collisions at RHIC
and the LHC.
The second way is to   increase  $\mu_B$ at low
temperatures, $T < T_c$, going from  Hadronic Matter to Quarkyonic Matter.
We suggest that this phase boundary is studied by heavy ion collisions at
moderate and low energies, such as those at the AGS, SIS, and at low energies 
at the SPS, and in the future at FAIR and NICA \cite{hohne}.

At a special value of the baryon
chemical potential and temperature, there is a triple point where
Hadronic Matter, the Quark-Gluon Plasma, and Quarkyonic Matter all coexist.
From experiment, Fig. \ref{line}, we estimate that this occurs for
\begin{equation}
\mu_B^{\rm triple \, pt} \approx 350-400~{\rm MeV} \; , \;
T^{\rm triple \, pt} \approx 150-160~{\rm MeV} \; .
\label{triple}
\end{equation}
This point is  presumably near where the linear and the flat temperature 
regime in Fig.~\ref{line} intersect.  We argue in the following how
this arises from a triple point.

In thermodynamics a triple point is the point in a phase diagram where three
lines of first order phase transitions meet.  A common example is where a gas,
liquid, and solid coexist at a given value of the pressure and temperature.
Since there are only first order phase transitions, no correlation length
diverges at the triple point.  For example, in the phase diagram of water, the
phases of vapor, water, and ice all coexist at the triple point.  There is
also a critical point in the phase diagram of water, but it is situated far
from the triple point, at much higher temperature and pressure.

The properties of strongly interacting matter at large density are
characterized by several order
parameters. One is the thermal Wilson or Polyakov loop, which
measures the degree of deconfinement reached.
This is strictly an order parameter in theories
without quarks, or in the limit of a large number of colors,
$N_c \rightarrow \infty$, if the number of flavors, $N_f$, is kept
fixed.
The second is the chiral condensate as an order parameter for chiral symmetry
breaking. Chiral symmetry is an exact symmetry when there are
two (or more) flavors of massless quarks.
The last is the density of baryons, which is an order parameter
even in the large $N_c$ limit, when $N_f$ grows with $N_c$
\cite{Hidaka:2008yy}.

Hadronic Matter is  confined and exhibits  chiral symmetry breaking.
It is technically difficult to define confinement for finite $N_c$  for
a finite number of quark flavors, since the potential that separates
quarks is never linear at large distances.
This argument has a precise meaning at zero $N_f$ or infinite $N_c$, or
for zero temperature.
Nevertheless, there should be a well defined region of low baryon density
and low temperature where the physical degrees of freedom are mesons.
This phase is also to a good approximation free of baryons since
their densities,
$n_B/M_B^3\sim e^{(\mu_B - M_B)/T} \le  10^{-2}$ for typical values of
$\mu_B$ and $T$ not too close to the phase boundary.

The Quark-Gluon Plasma is
deconfined with  restored chiral
symmetry, and has nonzero baryon number density
when $\mu_B \neq 0$.  It is composed of quarks and gluons, although
we note that lattice simulations indicate that the transition to
a deconfined state is rapid, but not discontinuous
\cite{ghk}.
This means that,  for a range of temperatures above $T_c$,
there is a ``semi'' Quark-Gluon Plasma,
in which the theory is only partially deconfined \cite{semi,pnjla,pnjlb}.
We neglect here the effects of the semi Quark-Gluon Plasma,
since lattice simulations \cite{lattice_review,ghk}
indicate that the energy density rises quickly to values close to the
ideal gas value near $T_c$, and this is the main quantity which will
concern us.  This is unlike the pressure, which does not approach
the ideal gas value until several times $T_c$, and for which the
semi Quark-Gluon Plasma is important.

Quarkyonic Matter is (approximately) confined, but has a large baryon
number density, and also a large energy density.
Whether chiral symmetry is restored in Quarkyonic Matter
is not yet fully understood.  Even at very high densities, there could
be residual chiral symmetry breaking from pairing effects near the
Fermi surface.  For the present discussion
it does not  matter when and how chiral
symmetry is restored in the Quarkyonic phase.

We remark that studies of the Sakai-Sugimoto model at nonzero quark density
serve as one realization of Quarkyonic matter \cite{sakai_sugimoto}.

At the outset  we concede that, in the strict thermodynamic sense,
the QCD  phase diagram might or might not
have a true triple point.  After all, the deconfining transition at
low $\mu_B$, and nonzero temperature, appears not to be of first order,
but a rapid crossover.  If the deconfinement transition remained a crossover
for all $\mu_B$ values, then the triple point would not be a true point,
since it would not connect matter separated by a true first order phase
transition.
It might happen that there is a second order critical end point  along the deconfinement line,
in which case the triple point might truly reflect three different phases connected by
first order phase transitions.

We do suggest that there is
a true triple point in the limit of an infinite number of colors
\cite{McLerran:2007qj,Hidaka:2008yy}.  In this limit, the deconfinement
transition is of first order \cite{teper}, and the Quarkyonic transition may
exist  \cite{Hidaka:2008yy}.  Thus the behavior for QCD may be reminiscent of
that for a large number of colors, and exhibit an {\it approximate} triple
point.

For the present discussion, it is not important whether the triple point is
exact, or only approximate.  What is important is that, in going from Hadronic
Matter to either the Quark-Gluon Plasma, or Quarkyonic Matter, there is a
large increase in the number of degrees of freedom.  Hadronic Matter is
dominated by Goldstone bosons.  In QCD, the hadronic phase has three types of
pions, and a relatively small amount of kaons; for $N_f$ flavors, there are
$N_f^2 -1 $ Goldstone bosons in the hadronic phase.  These Goldstone bosons
dominate bulk properties of the system for temperatures and quark chemical potentials,
$\mu_Q = \mu_B/N_c$ much smaller than $\Lambda_{QCD}$.  As one gets close to a transition temperature, massive degrees of freedom become important, eventually becoming so numerous
that a transition to a new phase of matter is induced.

As is well known, there are many more degrees of freedom in
the Quark-Gluon Plasma: for $N_c$ colors, there are
$2 (N_c^2 -1)$ bosonic and $4 N_c N_f$ fermionic,
or $16$ bosonic and  24(36) fermionic degrees of freedom in QCD with 2(3) favours.
We note that for the pressure and energy density,
ideal fermions contribute $7/8$ of a boson.

While Quarkyonic Matter is confined, the principal point of
Ref. \cite{McLerran:2007qj} is that the energy density, or equivalently
the number of degrees of freedom, can be counted as for
deconfined quarks.  While near the Fermi surface
the degrees of freedom are confined baryons, most of the energy
density is due to quarks, deep in the Fermi sea.
This is a coarse description of what is surely
a much more complicated reality.
If we assume that chiral symmetry remains broken in the Quarkyonic
phase, Quarkyonic Matter then has $N_f^2 -1$ bosonic,
and $2 N_c N_f$ fermionic, degrees of freedom.  The
number of fermionic degrees of freedom
is half that of the Quark-Gluon Plasma, since in
Quarkyonic Matter, only quarks, but not anti-quarks, contribute.
In QCD, there are 3(8) bosonic degrees of freedom, plus
 12(18) fermionic degrees of freedom for 2(3) flavours.
The number of degrees of freedom is smaller for Quarkyonic Matter
than for the  Quark-Gluon Plasma, but significantly larger than the number
of Goldstone degrees of freedom of Hadronic Matter.

Thus, we argue, that while there may be no true phase transitions from either
Quarkyonic Matter, or a Quark-Gluon Plasma, to Hadronic Matter, there is a
rapid decrease in the number of degrees of freedom and so in the energy 
density.
This rapid decrease could well cause the matter to decouple, and so define,
experimentally, the surfaces for chemical equilibrium.  This is approximately
true for the transition from the Quark-Gluon Plasma to the hadronic phase, as
observed at RHIC energies \cite{BraunMunzinger:2003zz}.

At RHIC energies, chemical freeze-out was shown  \cite{BraunMunzinger:2003zz}
to take place very close (within less than about 10 MeV) to the phase boundary,
driven by the rapid density change across the phase transition.
Further it was argued that freeze-out ends
when the system is fully hadronized, i.e. at low density in the hadronic
phase.  Were this not the case  \cite{knol}, one would also expect different freeze-out
parameters for each hadron species due to widely different hadronic cross
sections. This is not observed. We believe this argument to be
generic \cite{BraunMunzinger:2003zz}:
to ensure simultaneous (within a very small interval in temperature and
chemical potential) freeze-out of all hadrons, the freeze-out curve has
to be very close to a line with a rapid density change.
An immediate consequence of this
would be that the chemical freeze-out curve delineates phase boundaries, not
only for small values of $\mu_B$ but everywhere. But what provides the phase
boundary for large values of $\mu_B$, where the deconfinement transition seems
far away, at least if one follows the guidance from lattice QCD calculations?
As already indicated above we believe that the transition from Hadronic to
Quarkyonic Matter provides the missing link.

Across the Quarkyonic
line, we would expect that the transition takes place in a range of baryon
chemical potentials of order $\delta \mu_B \sim k_F^2/2M_B
 \sim  35$ MeV in width, for $\Lambda_{QCD} \sim 200$ MeV
as a typical baryonic mass scale in QCD and for $k_F=0.263$ MeV.  This width is parametrically of order
$1/N_c$ which  accounts for its anomalously small size compared to typical hadronic
energy scales.

\section{A Simple Hagedorn Model for the Quarkyonic Transition}

In this section, we explore a very simple model of the Quarkyonic transition. 
This model only counts the number of degrees of freedom of baryonic resonances,
and ignores effects due to the strong nucleonic interactions.  
It assumes that the resonance spectrum ``turns on'' in a very narrow window 
of $\mu_B$, as suggested  by the large $N_c$ arguments of the previous 
section.  Interaction effects should not therefore change the position of 
the phase boundary.  Nevertheless in realistic computations interactions 
should be taken into account, and a realistic spectrum of baryons should be 
used.
These modifications will be discussed in the next section.

Resonance formation is the dominant feature for mesonic interactions,
and the most detailed model of hadron dynamics, the dual resonance
model \cite{DRM}, in fact describes all scattering amplitudes in terms
of resonance poles in the different kinematic channels. The number of
states of mass $m$, the degeneracy $\rho(m)$, is found to increase as
\begin{equation}
\rho(m) \sim m^{-a} \exp\{2 \pi \sqrt{2\alpha'/3} m\}
\label{1}
\end{equation}
where $\alpha'\simeq 1$ GeV$^{-2}$ is the universal Regge resonance
slope and $a$ a positive constant \cite{DRM-deg}. A basic result in
the study of interacting systems is that if the interactions are
resonance-dominated, the system can be replaced by an ideal gas of
all possible resonances \cite{B-U,D-M-B}. The partition function
determining the thermodynamics of an ideal resonance gas \cite{H-W}
becomes
\begin{equation}
\ln Z(T,V) = {\rm const.}~ VT^{3/2}
\int_{m_0}^\infty dm~m^{(3/2)-a} e^{-m[(1/T) - (1/T_H)]},
\label{2}
\end{equation}
where $T_H^{-1}= 2 \pi \sqrt{2\alpha'/3}$. It is seen that this
partition function has a singular point at
$T_H  \simeq 190$  MeV,
indicating that the system cannot exist at higher temperatures.
Previous work assuming self-similar resonance formation, the
so-called Statistical Bootstrap Model \cite{Hage}, had also led to
an exponentially increasing level density, and for some time it was
assumed that $T_H$ was the ultimate temperature of matter. Subsequently
it was noted \cite{C-P} that $T_H$ marks a critical point, with a possible
new state of matter at $T>T_H$, which presumably is the Quark-Gluon
Plasma.

\medskip

An alternative approach is based on the intrinsic size of hadrons \cite{Pom}.
With increasing temperature, the hadron density increases, and - assuming
again a mesonic system - the individual constituents will overlap more and
more. At a certain density, the system will percolate, i.e., form a
connected network spanning the entire system. The spanning cluster
consists of overlapping mesons, so that it ceases to be meaningful
to speak of the existence of individual mesons within this cluster.
The density of mesons in the cluster is at the percolation point
approximately
\begin{equation}
n_p \simeq {1.2 \over V_0},
\label{4}
\end{equation}
where $V_0 \simeq (4 \pi/3) R_0^3$ and $R_0 \simeq 0.8$ fm. We can now
ask for the temperature at which an ideal resonance gas, with all resonances
having size $V_0$, attains this density. It is found to be \cite{C-R-S}
\begin{equation}
T_p \simeq 180~{\rm MeV},
\label{5}
\end{equation}
so that such geometric percolation considerations lead to a limit of
Hadronic Matter very much like that obtained from resonance dynamics.

\medskip

\begin{figure}[htbp]
\begin{center}
\resizebox*{!}{5.0cm}{\includegraphics{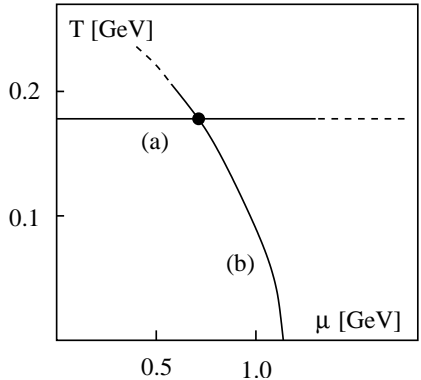}}
\end{center}
\caption{Limits of Hadronic Matter, (a) meson percolation or resonance
formation, (b) hard core baryon percolation.}
\label{diagram}
\end{figure}

The ``mesonic'' arguments used up to now continue to be valid also in the
presence of baryons, as long as the baryon density is well below the point
of dense packing; we will elaborate on this below. As a result, we
conclude that resonance formation or percolation lead to a temperature
limit $T_H$ approximately independent of the baryon
density\footnote{The existence of strange baryons does lead to a slight
decrease of $T_H(\mu)$ with baryochemical potential $\mu$ \cite{C-R-S};
we ignore this here for simplicity.}. Our ``phase diagram'' thus is so far
a straight horizontal line $T_H(\mu)=$ const. in the $T-\mu$ plane, as
shown in Fig.\ \ref{diagram}.

\medskip

The nature of the limit depends on the conceptual basis. An ideal resonance
gas with an exponentially growing mass spectrum results in a genuine
thermal critical line, corresponding to continuous transitions;
the associated critical exponents can be determined in terms of
the space dimension $d$ and the coefficient $a$ in Eq.\ (\ref{1})
\cite{H-W,HS74}. Percolation is in general a geometric critical phenomenon,
with singular behavior and corresponding critical exponents for
cluster variables. It does not imply singular behavior of the partition
function and could thus from a thermal point of view correspond to a
rapid cross-over \cite{HS01}.

\medskip

We now turn to the other extreme, dense baryonic matter at low
temperature. For baryochemical potential
$\mu \simeq 0$, the contribution of baryons/antibaryons and baryonic
resonances is relatively small, but with increasing baryon density, they
form an ever larger fraction of the species present in the medium, and
beyond some baryon density, they become the dominant constituents.
Finally, at vanishing temperature, the medium consists essentially of
nucleons.

\medskip

\begin{figure}[htbp]
\begin{center}
\resizebox*{!}{3.7cm}{\includegraphics{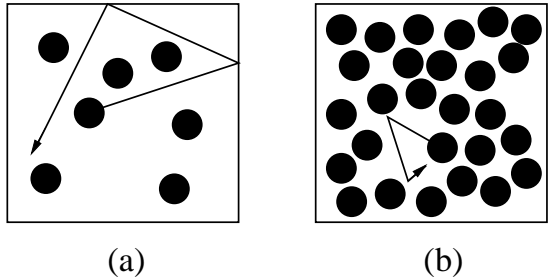}}
\end{center}
\caption{States of hard core baryons: full mobility (a), ``jammed'' (b)}
\label{hard}
\end{figure}

For vanishing or low baryon number density, when the interactions are
resonance dominated, the system could be described as an ideal gas of
all possible resonance species. At high baryon density, however, the
dominant interaction is non-resonant.
Nuclear forces are short-range and strongly attractive at distances of
about 1 fm; but for distances around 0.5 fm, they become strongly repulsive.
The former is what makes nuclei, the latter (together with Coulomb and Fermi
repulsion) prevents them from collapsing. The repulsion between a proton
and a neutron shows a purely baryonic ``hard-core'' effect and is connected
neither to Coulomb repulsion nor to Pauli blocking of nucleons. As a
consequence, the volume of a nucleus grows linearly with the sum of its
protons and neutrons. With increasing baryon density, the conceptual
basis of a resonance gas thus becomes less and less correct, so that
eventually one should encounter a regime of quite different nature.
At high baryon density, the most striking effect is the onset of a
``jamming'' of nucleons: the mobility of baryons in the medium becomes
strongly restricted by the presence of other baryons, leading to a jammed
state \cite{K-S}, as shown in Fig.\ \ref{hard}. The inverse mobility $s$
of a nucleon here plays the role of an order parameter: up to a certain
density, it is zero, and beyond this point, it remains finite.

\medskip

\medskip

Baryonic matter thus becomes again a medium of extensive hadrons of
radius $R_h$, but these now contain a hard core of a smaller radius
$R_{hc}< R_0$. The overlap of such hadrons in percolation studies is
thus restricted; nevertheless, the percolation onset can still be
determined \cite{Kratky}, and it is found \cite{C-R-S} that the density
of a spanning cluster now becomes
\begin{equation}
n_p^{hc} \simeq {2 \over V_0},
\label{6}
\end{equation}
assuming $R_{hc} = R_0/2$. With $R_0 = 0.8$ fm, this leads at $T=0$
to a limit of about 5.5 times standard nuclear density.
Requiring the baryon density (baryons minus antibaryons) in an ideal
resonance gas to attain this limit as function of $T$ and $\mu$ then
defines a critical curve based on baryon percolation. In the simplest
model,
\begin{equation}
\mu_p \simeq 1.12~{\rm GeV}
\label{7}
\end{equation}
becomes the limiting baryochemical potential $T=0$. The general curve
is included in Fig.\ \ref{diagram} \cite{C-R-S}.

\medskip

In the case of hard core percolation, a connection to thermodynamic
critical behaviour has also been discussed \cite{Kratky}. If a system
with hard core repulsion between its constituents is in addition
subject to a density-dependent negative background potential, first
order critical behaviour can appear, ending in a second order critical
point specified by the background potential strength and the hard
core volume.

\medskip
\begin{figure}[htbp]
\begin{center}
\resizebox*{!}{7.2cm}{\includegraphics{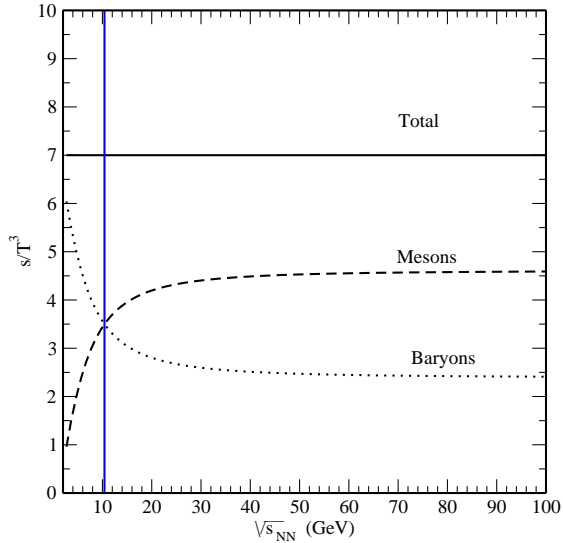}}
\end{center}
\caption{The
baryon number and mesonic contributions to the entropy density as a
function of center of mass energy for the collisions of heavy nuclei.  The
values of $\mu_B$ and $T$ used to make this plot arise from Statistical
Model parameterization of the chemical abundance of produced particles. \cite{corw1}}
\label{baryon}
\end{figure}

The interpretation of the situation illustrated in Fig.\ \ref{diagram}
allows different interesting possibilities. In Ref. \cite{C-R-S} it is
assumed that the state outside the Hadronic Matter region is a deconfined
Quark-Gluon Plasma. It is, however, also conceivable that below the
meson percolation/resonance curve confined mesonic states survive, while
baryons enter into the new phase. Such Quarkyonic Matter
\cite{McLerran:2007qj,Hidaka:2008yy}
is dealt with in detail in this work.

One can get some insight into the nature of the transition in the various
regions of $\mu_B$ and $T$ by plotting the entropy density inferred from
resonance model descriptions as a function of center of mass energy of the
collision, as shown in Fig. \ref{baryon}.  For low energies, below the
hypothetical critical
point, the matter is baryonic, consistent with a transition from Hadronic Matter
to Quarkyonic Matter.  For higher energies, it is largely mesonic matter,
and consistent with a transition from Hadronic Matter to a Quark-Gluon Plasma.
Turning this into $\mu_B$, one goes from a
region dominated by baryons at decoupling, when $\mu_B > 400$~MeV, to one
dominated by mesons at decoupling, for $\mu_B < 400$~MeV \cite{corw2}.

As a simple model that embodies some of the features discussed above,
we suggest that, for small values of the chemical potential,
$\mu_B < \mu_B^{\rm triple \, pt}$, the transition between a Hadronic phase,
and the Quark-Gluon Plasma, is controlled by a
single Hagedorn temperature\footnote{The precise 
relation between the QCD phase boundary and the Hagedorn temperature is 
not well understood at the moment. Our schematic construction, leading to 
Eq.~\ref{triple_point} below, implies asymptotically $T_H^M<T_H^B$.
Note, however, that, from the the presently-known hadron spectrum up to 2 GeV,
the effective $T_H$ for mesons appears to be larger than for 
baryons \cite{bro}.} 
for {\it mesons}, $T_H^M$
\cite{H-W,Hage,C-P,hagedorn,bro}.  Assuming that
this transition is controlled entirely by mesons, we obtain a line which
is independent of  $\mu_B$.  Of course we do not believe that this behavior
is exact, but it seems to be not a  bad approximation in QCD.
Numerical simulations of lattice QCD imply that
$T_c$ decreases very slowly, by only about $10\%$, for
$\mu_B$ from 0 to 400 MeV \cite{lattice_review,ghk}.
The $\mu_B$ independence of  $T_H^M$  is  also in accord
with arguments at large $N_c$ and small $N_f$, which imply that
the critical temperature is independent of the baryon chemical potential
(As we discuss below, this is true as
long as $\mu_B/N_c$ is of order one, and does not grow with $N_c$.)

We suggest that this horizontal  line intersects with a second line, which
is controlled by a
Hagedorn temperature for {\it baryons}, $T_H^B$.  If there is
such a Hagedorn temperature, the
density of states of baryons grows like
\begin{equation}
\rho_B(M_B) \sim \exp(+M_B/T_H^B) \; , \;
M_B\rightarrow \infty
\label{baryonic_hagedorn}
\end{equation}
We assume, as is typical for a Hagedorn spectrum, that this balances
against the usual
Boltzmann factor, $\exp((\mu_B - M_B)/T)$.  Then for a given
value of $\mu_B$, there is a phase
transition at a ``Quarkyonic''
temperature $T_{\rm Qk}$, which is $\mu_B$-dependent.
In the plane of $\mu_B$ and temperature, this dependence is just
a straight line:
\begin{equation}
T_{\rm Qk}(\mu_B) = \left( 1 - \frac{\mu_B}{M^0_B} \right) T^B_H \; .
\label{quark_temp}
\end{equation}
We have made a gross approximation in this formula, which is represented
by the parameter $M^0_B$.
The Hagedorn mass spectrum in Eq. (\ref{baryonic_hagedorn}) is
only valid asymptotically, as $M_B \rightarrow \infty$.  Thus
strictly speaking, the transition
temperature from a Hagedorn spectrum is independent of $\mu_B$.
(For this reason, in string models the Hagedorn temperature is
common to all particles, determined only by a single parameter,
which is the string tension \cite{Hage,hagedorn}.)
Instead, in Eq. (\ref{quark_temp}) we introduce a new parameter,
$M^0_B$, by hand.
This is meant to represent a {\it finite} mass scale at which a Hagedorn
spectrum appears.  Here $M^0_B$ is entirely a phenomenological
parameter, meant to illustrate how the transition temperature   $T_{\rm Qk}$ to
Quarkyonic Matter might depend upon $\mu_B$.  Clearly
$M_0^B$ cannot be less than the mass of the lightest baryon; it could
well be much larger.

As one decreases $\mu_B$, eventually there will be a temperature at
which this line crosses that for deconfinement.  We assume that, when
this happens, the line for the Quarkyonic transition ends and
that the transition to a Quark-Gluon Plasma, which has a much larger
energy density, dominates.  The point at which these two lines cross
defines the position of a triple point:
\begin{equation}
T^{\rm triple \, pt} = T^M_H
= \left( 1 - \frac{\mu_B^{\rm triple \, pt} }{M^0_B} \right) T^B_H \; .
\label{triple_point}
\end{equation}
We stress that our approximations are {\it very} crude,
and are only meant to illustrate how
a triple point {\it might} arise.

The transition temperature line of Eq. (\ref{quark_temp}) intersects the
axis of $T = 0$ when $\mu_B = M^0_B$.
This formula does not apply at arbitrarily low temperatures,
however.  A Quarkyonic phase is defined to be one in which both the
baryon and energy densities are large, but at small temperatures, when
$\mu_B$ is close to the nucleon mass, one probes not Quarkyonic, but dilute
nuclear matter.  At large $N_c$, though,
the region in which nuclear matter is dilute is a narrow
window in $\mu_B$ \cite{McLerran:2007qj}.
This suggests that $M_B^0$ is near the nucleon mass;
fitting to Fig. \ref{line} gives $M^0_B \approx 1$~GeV.
The value of the baryonic
Hagedorn temperature can be read off from where $T_{Qk}(\mu_B)$ intersects
the axis of $\mu_B = 0$.  Again from Fig. \ref{line}, this gives
$T_H^B \approx 250$~MeV.

These values for mesonic and baryonic
Hagedorn temperatures should only be taken as illustrative.  Even at
$\mu_B = 0$, experiment gives us the results at chemical freeze-out.
This value is certainly lower than the temperature for the true
transition (or crossover), and is lower still than that for the Hagedorn
temperature.  One might, however, expect that these values are close to
one another.  This is indicated by results from the lattice in
pure gauge theories \cite{lattice_hagedorn}.

The limit of a large number of colors shows that
the introduction of the parameter $M^0_B$ is not quite as contrived as
might first appear.  In the limit of large $N_c$ and small $N_f$
the transition from a hadronic to a Quarkyonic phase is a straight
line along $\mu_B = m_N$, where $m_N$ is the mass of the lightest baryon
(up to small effects from nuclear binding) \cite{McLerran:2007qj}.
This is just the usual mass threshold for a chemical potential.

In the limit in which both $N_c$ and $N_f$ are large,
one cannot speak of deconfinement rigorously, and there is
only a phase transition for the condensation of baryons,
which is a straight line in the $\mu_B - T$ plane
\cite{Hidaka:2008yy}.  This is because the density of
states, for even the lowest baryon multiplet, grows exponentially.
Note that this
is analogous, but not identical, to a Hagedorn temperature, since
the exponential growth is for the lightest multiplet, and not for
asymptotically large masses.  There are several effects which will
act to modify this naive prediction, however.
First, even in the Hadronic phase, baryons interact strongly
with the numerous mesons.  This will modify the baryon mass, and so shift
the threshold at which they condense.
Second, baryon baryon interactions are strong.
In ordinary nuclear matter it is well known that baryons have
a large hard core repulsion between them, and this surely persists
when both $N_c$ and $N_f$ are large.  Such a hard core repulsion between
baryons will act to cut off the singularity in the free energy,
which otherwise would be generated by an exponential growth in the
degeneracy of states.

Of course in QCD the degeneracy of the lowest baryons does not
grow exponentially.
But $M^0_B$ can then be viewed as a way of
characterizing when the growth of baryonic states starts to take off.
For example, this could be estimated more accurately in resonance gas
models.  Consider, alternately, the result of \cite{Hidaka:2008yy}:
\begin{equation}
T_{\rm Qk}(\mu_B) = \frac{M_B - \mu_B}{\log(N_{\rm deg}) } \;
\label{largeNf}
\end{equation}
In this equation, $N_{\rm deg}$ is the number of approximately
degenerate baryonic states.
Extrapolating the formula for large $N_c$ and $N_f$ down to small values,
for three colors and two flavors, $T_{\rm Qk} \approx 160$~MeV;
for three flavors, $140$~MeV  for $\mu_B\simeq 400 $ MeV \cite{Hidaka:2008yy}.
In QCD we can also estimate $N_{\rm deg}$ directly.
Including nucleons and
the $\Delta$ resonance, $N_{\rm deg} = 20$; including
all strange baryons, $N_{\rm deg} = 56$.
Further, by including higher resonances, in Eq. (\ref{largeNf}) we should
take not the nucleon mass for $M_B$, but some heavier state, which
can then be used to define $M^0_B$.
By fitting the data in Fig. \ref{line}, with $M^0_B \sim 1$~GeV,
one finds that $\log(N_{\rm deg}) \approx 2-3$, which is
not too far from the extrapolation from large $N_c$.  Of course, the precise
tradeoff between the increasing masses of various states and their abundances
is a tricky issue.  It is not clear how much to include of the flavor and spin
excitations of the lowest mass nucleon states, and hence the uncertainty.
This may be addressed more directly within Statistical Model computations.

A natural question is what happens to the two phase
transition lines beyond the triple point.
Consider first the transition between the Hadronic  phase
and the Quark-Gluon Plasma, to the right of the triple point
at approximately constant temperature, with
$\mu_B > \mu_B^{\rm triple \, pt}$.
At large $N_c$, this line is of first order, and
remains a boundary for a true phase transition.
The lattice QCD results show that the rapid rise in the energy
density is relatively independent of $\mu_B$; thus we suggest that this
line delineates an approximate phase transition for
$\mu_B > \mu_B^{\rm triple \, pt}$.
At large $N_c$,  when $\mu_B/M_N \sim N_c^{1/2}$,
eventually deconfinement is washed out by the quarks, and there is
a critical endpoint for deconfinement.  (This is the value of chemical
potential where the Debye screening length becomes less than the confinement
size scale.)  In QCD, since there isn't a first order transition to
deconfinement, we expect that eventually
the large increase in the energy density, seen in a narrow region in
temperature, is just washed out by the contribution of dense quarks.

It is also possible to consider continuing the phase boundary for the
Quarkyonic phase at chemical potentials
below the triple point, that is, for $\mu_B < \mu_B^{\rm triple\, pt}$.
One might imagine that there is then a line
for the Quarkyonic transition {\it above} that for deconfinement,
with $T_{\rm Qk} > T_c$ when $\mu_B < \mu_B^{\rm triple\, pt}$.
Even at large $N_c$, such a line of Quarkyonic ``transitions'' can only
reflect the properties of
some metastable state in the (semi-)Quark-Gluon Plasma.
Numerical simulations on the lattice do find that like the energy density,
quark number susceptibilities approach the ideal gas values very near $T_c$,
by $20\%$ above $T_c$  \cite{lattice_review}.  Thus perhaps this
change in the quark number susceptibilities
reflects the remnants of the Quarkyonic ``transition'' in the deconfined phase.

We conclude by discussing the relationship to the chiral phase transition.
It is possible that the triple point coincides with a critical
end point for a line of first order (chiral) transition \cite{Stephanov}.
However, as we noted above, the experimentally observed properties of
the triple point do not seem indicative of a critical point.  It is
possible that QCD  matter behaves similar to  water, with a critical end point for the chiral
transition which is distinct from the triple point.  If so, it probably
exists in the Quarkyonic phase.  It is also possible that
there is no well defined chiral transition; {\it i.e.}, that because
of the nonzero quark masses, the chiral transition is only a crossover.
There would then be no critical end point for the chiral phase transition.

\section{Strangeness along the phase boundary}

We have already discussed and shown in Fig.~\ref{fig_k2pi} that there are
abrupt changes in the abundances of various ratios of strange to
non-strange particles at $\sqrt s$ around 10 GeV.\footnote{Note that the energy  axis
in Figs. \ref{fig_k2pi}
is in logarithmic scale so the variation at higher
energies is indeed quite slow.}
The reason for such
behaviour  may be linked with the appearance of the Quarkyonic phase.  Along the Quarkyonic line, the
temperature changes substantially.  The fraction of strange particles should increase as the
temperature increases.  Along the Quark-Gluon Plasma line, the temperature
is constant  and we
 expect the strange quark relative abundance to be roughly unchanged.
When these two boundaries meet, we would expect a change  in the
strange quark density near the triple point.  This is most easily seen
approaching the triple point along the Quarkyonic curve since as one
approaches the deconfinement transition, there should be a rapid increase in
the energy density, favoring a higher relative abundance of strange quarks.
The strange quark relative abundance  increases rapidly
as one approaches the triple point, but then slowly decreases beyond it
due to  the decreasing $\mu_B$ at the almost constant  temperature.

\begin{figure}
\resizebox{0.985\textwidth}{!}{%
\includegraphics*[width=7.5cm]{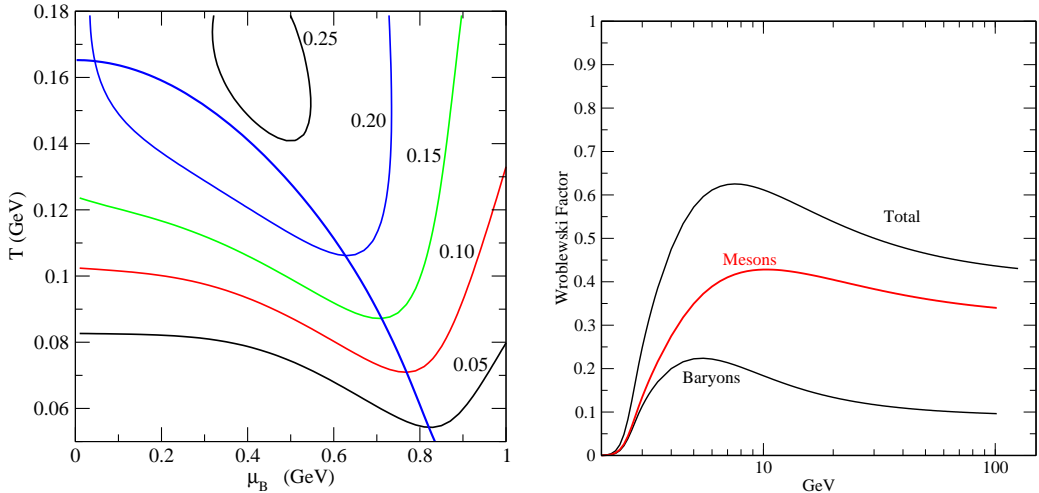}\hspace{5mm}
\includegraphics*[width=7.2cm]{wroblewski}}
\caption{The left hand figure:
Contours of constant values of the  $K^+/\pi^+$ ratio in the $T - \mu_B$ p
lane \cite{ho}. The line is  the $E/N $ line from Fig.~\ref{line}. 
The right hand figure: The Wr\'oblewski factor as a function of energy with 
separate contributions of mesons and baryons.
}\label{wroblewski}
\end{figure}

Some of the strange to non-strange particle ratios are very sensitive to small
variations in $T$ and/or $\mu_B$ as demonstrated in Fig. \ref{wroblewski}-left
showing the $K^+/\pi^+$ ratio as contour lines in the $T-\mu_B$ plane
\cite{ho}.  If in the region of the triple point the freeze-out would happen at
somewhat higher temperatures then especially this ratio will increase.  Fig.
\ref{wroblewski}-left illustrates that in the Statistical Model the $K^+/\pi^+$
ratio can never exceed the value of $\sim 0.25$.

 While different strange to non-strange particle ratios exhibit different
 trends, the relative  strangeness content quantified  by the Wr\'oblewski factor\footnote{The Wr\'oblewski factor,  $2N_{s\overline{s}}/(N_{u\overline{u} }+
N_{d\overline{d}})$,  determines the
relative abundance of the initially produced  strange to light quarks multiplicities.},
 similarly   as $K^+/\pi^+$, exhibits a well pronounced  peak   as seen   in Fig. \ref{wroblewski}-right.

The peak in the strangeness abundance naturally arises in the Statistical Model
due to the presence of the phase boundaries between QGP-Quarkyonic Matter and
Hadronic Matter, for the reasons stated above.  It is nevertheless an indirect
measure of the singularity associated with a triple point.  If the triple point
region is somewhat spread out, we might expect that the peaks in various
particle ratios might not appear at the same point.  If the critical point
region is very narrow, there should be approximately discontinuous behavior at
the critical point, but this does not necessarily imply a maximum in ratios of
strange to non-strange particles near to the critical point. The relationship
between strangeness abundance, the triple point, and experimental data is
certainly worth more detailed and precise experimental and theoretical study.

\section{Quarkyonic Matter and Chiral Symmetry Breaking}

So far chiral symmetry played  little role in our argument;  the hadron
resonance gas description assumes no explicit modification on the
hadron masses in a hot and dense medium.  In principle, however, it
would be conceivable to anticipate a substantial change in the hadron
spectrum depending on whether chiral symmetry is (partially) restored
or not.  There are in fact several theoretical and experimental
indications that chiral symmetry is affected in a medium~\cite{Hayano:2008vn}.
For instance,
the leading-order of the virial expansion suggests that the chiral
condensate receives, at normal nuclear density, a $30$-$40\%$ reduction,
which is shifted  back by around $10\%$ by higher order corrections in
the in-medium chiral perturbation theory~\cite{Kaiser:2008qu}.  Of
course, the chiral condensate itself is not a direct experimental
observable, but useful information is available from the spectroscopy
of  deeply-bound pionic atoms and the experimentally deduced
in-medium pion decay constant at normal nuclear density is reduced by
36\% compared to its vacuum value~\cite{Suzuki:2002ae}.

Although  chiral perturbation theory gives a fairly
model independent statement on chiral properties at finite temperature
and baryon density, forming a productive research area together with
experimental measurements, its validity is strictly limited to
low-energy regimes.  As one tries to go  beyond   low-energy regimes to
explore the phase diagram of strongly interacting QCD matter, it is
extremely difficult to make any statement in a model-independent way.
It is notable that the in-medium condensate strongly relies on the pion
mass coming from two-pion exchange correlations with virtual $\Delta(1232)$
excitations which stabilizes the dropping of the condensate for physical
pion mass with increasing density~\cite{Kaiser:2008qu}.
Particularly, a deviation from the result in the linear density approximation
is remarkable for symmetric nuclear matter.
This might indicate that the chiral symmetry restoration  would
take place at much higher density as compared to the critical density
given in the mean-field models.
This also could suggest that in-medium
correlations might weaken a phase transition  and eventually
a first-order phase transition might disappear from the phase diagram.

In the context of the chiral phase transition the
most frequently used models are the Nambu--Jona-Lasinio (NJL) model
and the quark meson (QM) model, which are sometimes improved by the
introduction of partial gauge degrees of freedom, namely the
Polyakov loop, and promoted to the PNJL and PQM \cite{schaefer1} models, 
respectively.
Crucial points in this sort of model treatment are that a description
in terms of quasi-quarks is assumed  and the effect of the confinement
is totally neglected.
Such chiral quark models as well as another non-perturbative approach
using the Schwinger-Dyson equation~\cite{SDE}
favor a first-order chiral phase transition at high density.
This suggests a termination point of the first-order
phase boundary, which defines a critical endpoint (or often called the
QCD critical point).
Results from finite-density lattice simulations are far from conclusive 
yet and thus the existence of the critical endpoint is still under
extensive dispute.
In a description in terms of quarks, the driving force to induce a chiral
transition is the density contribution to the pressure.  Therefore, a
chiral phase transition in this region of low-$T$ and high-$\mu_B$ is
always acompanied by a significant jump in the quark number density.
So, if the correct degrees of freedom are quarks rather than baryons,
the quarkyonic transition is naturally close to the chiral phase
transition~\cite{Fukushima:2008wg}.
In these kinds of models there is a general tendency that the critical
endpoint is found not far from the triple point region.  This is
because the quarkyonic transition boundary tends to stay along the
chiral phase transition where the quark number density jumps
discontinuously or increases rapidly.
One must, however, bear in mind that the above-mentioned model
indications on the critical endpoint are strongly dependent on
neglected effects.
These include the unknown model
parameters and their dependence on $T$ and $\mu_B$ in the NJL and QM
models, other possibilities of ground states such as the color
superconducting phase, inhomogeneous states like the crystalline color
superconducting phase and chiral density
wave~\cite{Nakano:2004cd,Nickel:2009wj},  an unconventional pattern of
chiral symmetry breaking~\cite{hst}, etc.  Of course, in addition to them, the
baryon degrees of freedom may change the whole picture completely.

Let us consider how the quark model results are affected by the baryons.
To this end, we
must consider how the baryon belongs to the representation of chiral
symmetry.  There are two possible assignments;  one is just the same
as the quark field which is called a naive assignment, and another is
the so-called mirror assignment~\cite{Detar:1988kn}.
The important point is that one can
construct a mass term which is chiral invariant in the case of the
mirror assignment.  This means that the baryons need not be lighter
associated with chiral restoration in this case, so there is no jump
in the baryon number density across the chiral phase transition if
any.
Thus the baryon number density need not necessarily  exhibit a clear
indication of either Quarkyonic or chiral transitions.
On the other hand, if the assignment is naive, the situation
becomes more or less similar to what we have seen for the quark model
studies at the qualitative level.

Finally we shall comment on what the chiral model suggests for the
triple point.
In the PNJL model at small chemical potential ($\mu_q\ll T\sim T_c$),
the behavior of
the Polyakov loop as a function of $\mu_q$ and $T$ has a deviation
from that of the chiral condensate.  Such a possibility of unlocking
of the deconfinement and chiral transitions was already pointed out
in the first paper on the PNJL model~\cite{pnjla}.
Together with the fact that the quark number density has a strong
correlation with the chiral condensate in the quark-based model, this
observation of separate deconfinement and chiral crossovers may well
suggest that there appears a triple point region on the phase
diagram.
At least within the uncertainty of the
model which can easily move the critical endpoint, it seems that the
appearance of the triple point is a robust feature of the model
output.
It is notable that the anomaly matching condition may
well imply that the chiral phase transition takes place later
than the deconfinement~\cite{Baym:1982sn}, but strictly speaking, because of
the violation of Lorentz symmetry in the presence of matter,
the same argument as in vacuum cannot be directly applied to
constrain the ordering of the phase transitions~\cite{anomalymatching}.

While the arguments that chiral symmetry should be approximately restored in
the high
baryon density region are strong, the arguments that it is completely
restored are less so.  It might turn out for example that chiral symmetry
remains broken in the Quarkyonic phase due to non perturbative
effects at the Fermi  surface.  Such
effects would be proportional  to powers of $\Lambda_{\rm QCD}/T$,
and would be small but nevertheless non-negligible.
Presumably these effects would disappear when confinement also disappears.
For example, effects at the Fermi surface might make
the chiral condensate of order $\Lambda_{\rm QCD}^3$.
 While this would be small compared to the baryon number density, $\mu_Q^3$,
and might be ignored for many purposes,
its magnitude would be parametrically unchanged from its value in the confined phase.
Although we have very little to say which is strongly compelling about the nature of chiral symmetry
breaking and its relation to Quarkyonic Matter, the questions that arise are of fundamental interest
for our understanding of the nature of mass generation in QCD.  As such, these issues must eventually be understood in an absolutely compelling and simple way.

\section{The Triple  and Critical Point within an Effective Theory}

The possible relation between the chiral and deconfining phase transitions, discussed above, can be more transparent when referring to properties of an effective  Lagrangian \cite{mocsy}.

Consider the interaction of the (renormalized) Polyakov loop, $\ell$,
which is the trace of the renomalized Wilson line, $L$,
$\ell = {\rm tr} \, L/N_c$.  The Polyakov loop
couples to the chiral field, $\Phi$, as
\begin{equation}
L_{\rm eff} = c_1 \; \ell \; {\rm tr} \; \Phi^\dagger \, \Phi \; .
\label{eff1}
\end{equation}
This term is chirally invariant.  It is not invariant under the global
$Z(N_c)$ symmetry of the pure glue theory, under which
$\ell \rightarrow \exp(2 \pi i/N_c) \, \ell$, but this symmetry is broken
by the presence of quarks.  While the Polyakov loop $\ell$ is dimensionless,
for the purposes of power counting, let us assume that like ordinary
scalar fields, it has dimensions of mass.  This implies that
the coupling $c_1$ has dimensions of mass.  It is the
dominant coupling of the Polyakov loop to quarks.

The observation of Ref. \cite{mocsy} is that the sign of $c_1$ controls
how the chiral and deconfining transitions are related.  Assume
that chiral symmetry is broken in the vacuum, so
the expectation value of ${\rm tr} \, \Phi^\dagger \Phi$ is nonzero.
If $c_1$ is positive, $L_{\rm eff}$ is positive, so that this term
resists the Polyakov loop from developing an expectation value until
chiral symmetry is restored.  That is, positive $c_1$ links the
deconfining and chiral symmetry phase transitions together.
Conversely, if $c_1$ is negative, the transitions tend to repel
one another.

Clearly a special point occurs when $c_1$ vanishes.  At this point, it
is natural for the deconfining and chiral phase transitions to split apart
from one another.  We suggest, then, that the chiral phase transition {\it may}
split from the deconfining line at the triple point.  This assumes that
the triple point also coincides with the critical endpoint for the
chiral phase transition \cite{Stephanov}.

About the triple point, it is then natural to ask what the next
leading term is.  There are many such terms.  One involves the mass matrix
of the chiral fields, $M$, which is proportional to the current quark
masses:
\begin{equation}
L_{\rm eff}'
= c_2 \; \ell \; {\rm tr} \; M \Phi =
c_2 \; \ell \; \left( m_\pi^2 \pi^2 + m_K^2 K^2 + \ldots \right)/f_\pi \; .
\label{eff2}
\end{equation}
Like $c_1$, it has dimensions of mass.  It is chirally suppressed, however,
and so is less important except when $c_1$ is small.

Assuming that $c_1$ vanishes at the triple point, the coupling $c_2$
dominates in the region where $c_1 \approx 0$.  In this region, the
coupling is reversed from the usual expectation: in particular, the coupling
to the heavier Goldstone bosons, such as kaons, is {\it larger} than that
to the lighter Goldstone bosons, the pions.

The phenomenological implications of this term are interesting for
experiment.  About the point where $c_1$ vanishes, even though they
are heavier, the coupling of kaons to the Polyakov loop is larger,
by a factor of $m_K^2/m_\pi^2 \sim 13$.  This might explain the
enhancement of strangeness observed at the triple point.

\section{Summary and Conclusions}

In this work, we have presented an interpretation of  the  experimental data
on particle production obtained in heavy ion collisions  form SIS up to RHIC
energies in the context of a new  structure of the  QCD phase diagram with
Quarkyonic Matter.
 We have shown that  by considering Quarkyonic Matter, the two regimes of 
chemical freeze-out with meson and baryon dominance  observed phenomenologically
can be understood as arising from a triple point where Hadronic Matter, the
Quark-Gluon Plasma, and Quarkyonic Matter all coexist.  This triple point is
located where the freeze-out temperature is reaching its limiting value and were
different strange to non-strange particle ratios exhibit  non-monotonic behavior.

We have presented a set of qualitative and semi-quantitative arguments that the 
observed statistical properties of experimental data are naturally explained when
assuming the existence of   Quarkyonic Matter and of a  triple point in the 
QCD phase diagram. We have also discussed in the context of different models 
the possible role of the chiral symmetry restoration and the interplay between 
triple and critical points.

Our findings and interpretation can  be justified   and/or  verified in the near
future with more data expected from  ultrarelativistic heavy ion collisions.
New data are soon to be available  from the RHIC low energy runs and from the
CERN NA61 experiment which aim at a scan of energy and system size dependence
in the vicinity of the  triple point discussed here.
While these experiments will, most likely,  pierce into the triple point region
coming from higher energies, where the phase transition is a crossover between 
Hadronic Matter and Quark Plasma, the dedicated exploration of the phase border 
between the hadronic and the Quarkyonic Matter will be a task for the future
experiments with highly compressed baryonic  matter such as CBM@FAIR Darmstadt
and NICA@JINR Dubna.
If chemical equilibration will be substantiated with more precise data 
at those energies our quarkyonic phase boundary argument will gain a dramatic
support as a viable explanation for equilibration.

\section*{Acknowledgments}

We gratefully acknowledge insightful  comments from  Jean-Paul Blaizot and
Christof Wetterich.
The research of D. Blaschke is supported by the
Polish Ministry of Science and Higher Education (MNiSW) under
grants No. N N 202 0953 33 and No. N N 202 2318 37,
and by the Russian Fund for Fundamental Investigations
under grant No. 08-02-01003-a.
The research of R. Pisarski and L. McLerran is supported under DOE Contract
No. DE-AC02-98CH10886. R. Pisarski and K. Redlich thank the Alexander
von Humboldt Foundation (AvH) for their support;
K. Redlich also thanks the Polish Ministry of Science (MNiSW) for their
support. K. Fukushima  is supported by Japanese MEXT grant No.\ 20740134. The work of C. Sasaki was supported in part
by the DFG cluster of excellence ``Origin and Structure of the
Universe''.


\end{document}